\documentclass{ws-procs11x85}
\usepackage{balance}

\makeindex

\newcommand{\met}{\hbox{E\kern-0.5em\lower-0.1ex\hbox{/}}_T}

\begin{document}

\title{Bias-Free Estimation in  Multicomponent Maximum Likelihood Fits
with Component-Dependent Templates}

\author{P. Catastini}

\address{I.N.F.N.-Sezione di Pisa, Universit\'a di Siena,
 Italy\\E-mail: pierluigi.catastini@pi.infn.it}

\author{G. Punzi}

\address{I.N.F.N.-Sezione di Pisa, Scuola Normale Superiore,
Pisa, Italy\\E-mail: giovanni.punzi@pi.infn.it}


\twocolumn[\maketitle\begin{abstract} The possibility of strong biases
in a  multicomponent Maximum Likelihood  fits with component-dependent
templates has been demonstrated in  some toy problems. We discuss here
in  detail a  problem of  practical interest,  particle identification
based  on time-of-flight or  $dE/dx$ information. We show   that large
biases  can occur  in estimating  particle  fractions in  a sample  if
differences between the momentum spectra of particles are ignored, and
we present a more  robust fit technique, allowing bias-free estimation
even when the particle spectra in the sample are unknown.
\end{abstract}]

\bodymatter

\baselineskip=13.07pt
\section{Introduction}

It has been shown in  some toy problems\cite{punzi} that strong biases
may  occur in  a multicomponent  Maximum Likelihood  fit  whenever the
templates, i.e. the functions, used  to parameterize the probability  
distributions used in the fit are not fixed but depend on event 
observables.  An interesting example of such a problem  in the practice 
of experimental High Energy  Physics is the  statistical separation  of 
different  kinds of particles   on  the  basis   of  limited--precision   
measurements  of particle--dependent quantities,  like Time--of--Flight 
or  energy loss ($dE/dx$).

\section{Particle Fractions estimation}

Consider a sample of particles generated by a certain physical process
in our experiment. We know that the given sample is a mixture of known
particle  types,  for  example  $Pions$, $Kaons$  and  $Protons$,  but
unfortunately we  don't know the fractions of  each type, respectively
indicated  by  $f_{\pi}$, $f_{K}$,  $f_{P}$.   Let's  assume that  our
experimental  apparatus  includes  a Particle  Identification  ($PID$)
device,   providing   the  measurement   of   some  quantities   whose
distribution  depends   on  the  particle  type.    Using  this  $PID$
information  we want to  estimate $f_{\pi}$,  $f_{K}$ and  $f_{P}$, by
means of an Unbinned Maximum Likelihood fit of our data sample.

The above problem  is very common in particle  physics, for example it
occurs   in   separating   different    decay   modes   of   a   given
particle\cite{ichep04} (same final state multiplicity and topology but
different final state particle  types), in studies of fragmentation of
heavy   quarks\cite{tev},  or  in   optimizing  the   performances  of
algorithms for tagging the flavor of $B$ mesons\cite{tev}.

\begin{figure}
\begin{center}
\psfig{file=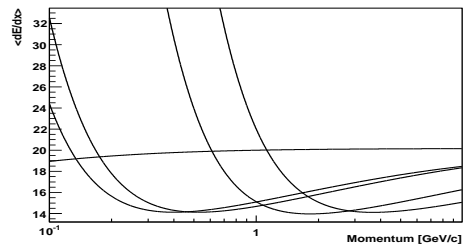,width=2.7in,height=1.4in}
\caption{The mean value  of the energy loss of  charged particles as a
function of the momentum in a typical experiment.}
\label{fig:dedx}
\end{center}
\end{figure}

We will  consider two common methods for  particle identification: one
is based on the measurement of energy loss of charged particles due to
the ionization of  a gas or of a semiconductor  (often the same device
used to measure particle momentum), the so called $dE/dx$ measurement;
the other is based on the measurement of the Time-of-Flight ($TOF$) of
the  particle.  A common  feature of  PID devices  based on  the above
principles is that the separation power between different particles is
not a  constant, but  strongly depends on  the momentum of  the given,
unknown, particle.   A clear example of  this feature is  shown in Fig
\ref{fig:dedx} where the $dE/dx$  mean response of different particles
is plotted as a function of momentum in the drift chamber of a typical
High-Energy  Physics  experiment. Assuming that the resolution of  the
measurement is constant, the  separation power dramatical changes in a
short momentum range.
As a  consequence of  the dependence  of the mean  value of  the $PID$
response on the particle  momentum, the templates describing the $PID$
variable's $p.d.f.$  are not fixed  but depend, on  an event-by-event
base, on the momentum of the particle: we clearly are in the situation
described in\cite{punzi}  where the templates  of the fit depend  on a
component of the fit itself.
\subsection{The Likelihood expression}

Consider,  for simplicity, only  the $PID$  information provided  by a
$dE/dx$ measurement.  Our observables are then the $dE/dx$ ($pid$) and
the  momentum of  the track  ($mom$).We  will indicate  as $type$  the
particular particle hypothesis.  Unfortunately, we cannot simply write
the Likelihood function as:
\begin{equation}
L(f_{j})=\prod\limits_{i}(\sum\limits_{j=\pi,K,P}  f_{j}  P(pid_{i}  |
mom_{i}, type_{j})).
\label{eq:likewrong}
\end{equation}
Using  expression (\ref{eq:likewrong})  may   give  a  strongly biased
result  if  our  additional  variable,  the  momentum,  has  different
distributions depending  on the particle  type (see next  section). As
discussed   in\cite{punzi},   whenever  the   templates   used  in   a
multi--component fit  depend on  additional observables, to  avoid the
bias  it  is  necessary   to  use  the  correct,  complete  Likelihood
expression,  including the explicit  distributions of  all observables
for all  classes of events.   In our case,  the above implies  that we
need to include  in our Likelihood the momentum  distributions of each
particle  type.   We  should  also  notice  that   in  practice  those
distributions are almost always different.

We then write the correct Likelihood function as:
\begin{eqnarray}
\label{eq:likeright}
L(f_{j})=\prod\limits_{i}(\sum\limits_{j=\pi,K,P}    f_{j}   P(pid_{i},
mom_{i}                                                              |
type_{j})\\\nonumber{=\prod\limits_{i}(\sum\limits_{j=\pi,K,P}    f_{j}
P(pid_{i}   |   mom_{i},   type_{j})}\\\nonumber{\times  P(mom_{i}   |
type_{j})),}
\end{eqnarray}
with the condition:
\begin{equation}
\label{eq:condition}
\sum\limits_{j=\pi,K,P} f_{j} = 1.
\end{equation}
\section{A toy study}

\begin{figure}
\begin{center}
\psfig{file=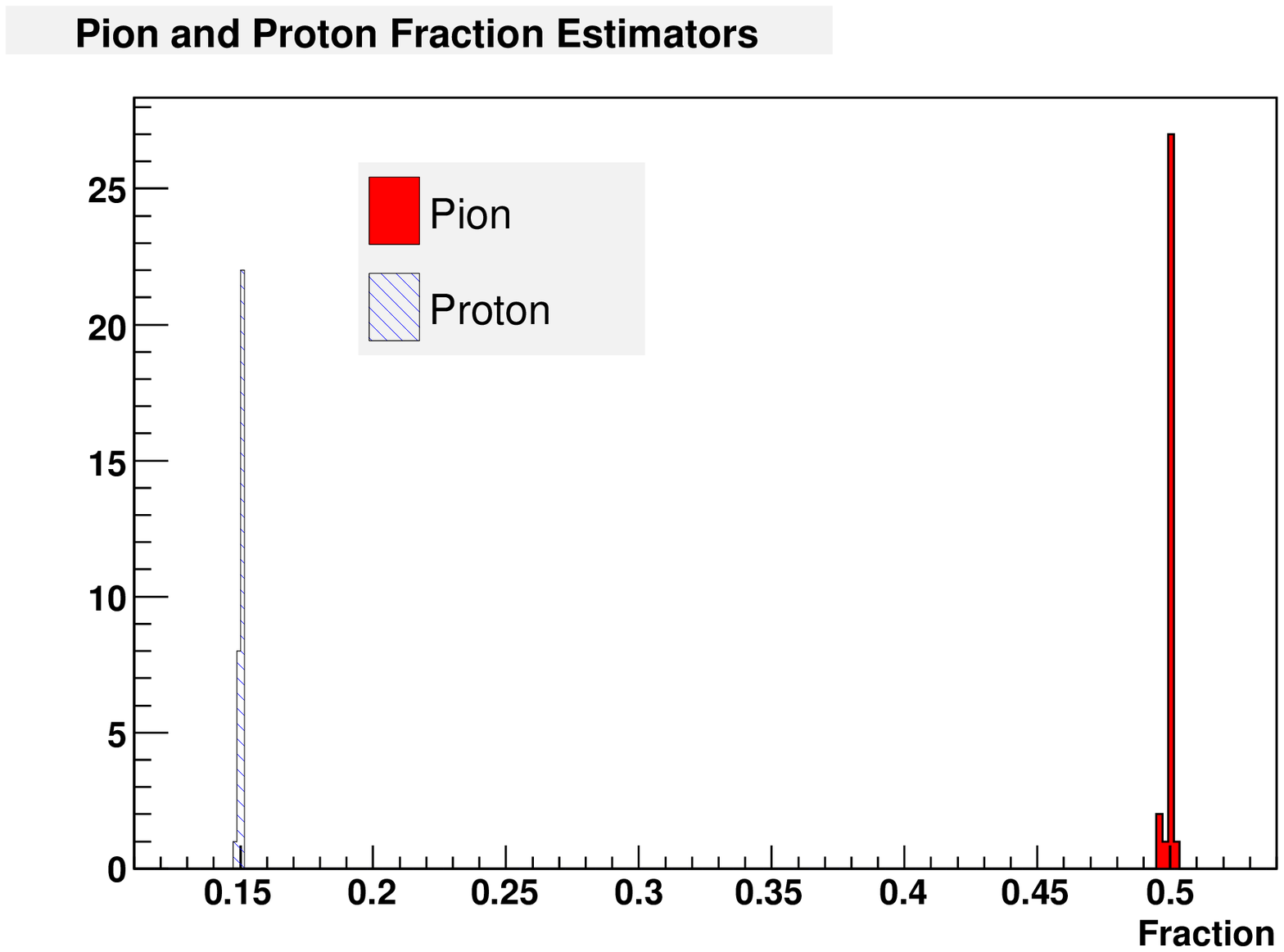,width=2.7in,height=1.4in}
\end{center}
\begin{center}
\psfig{file=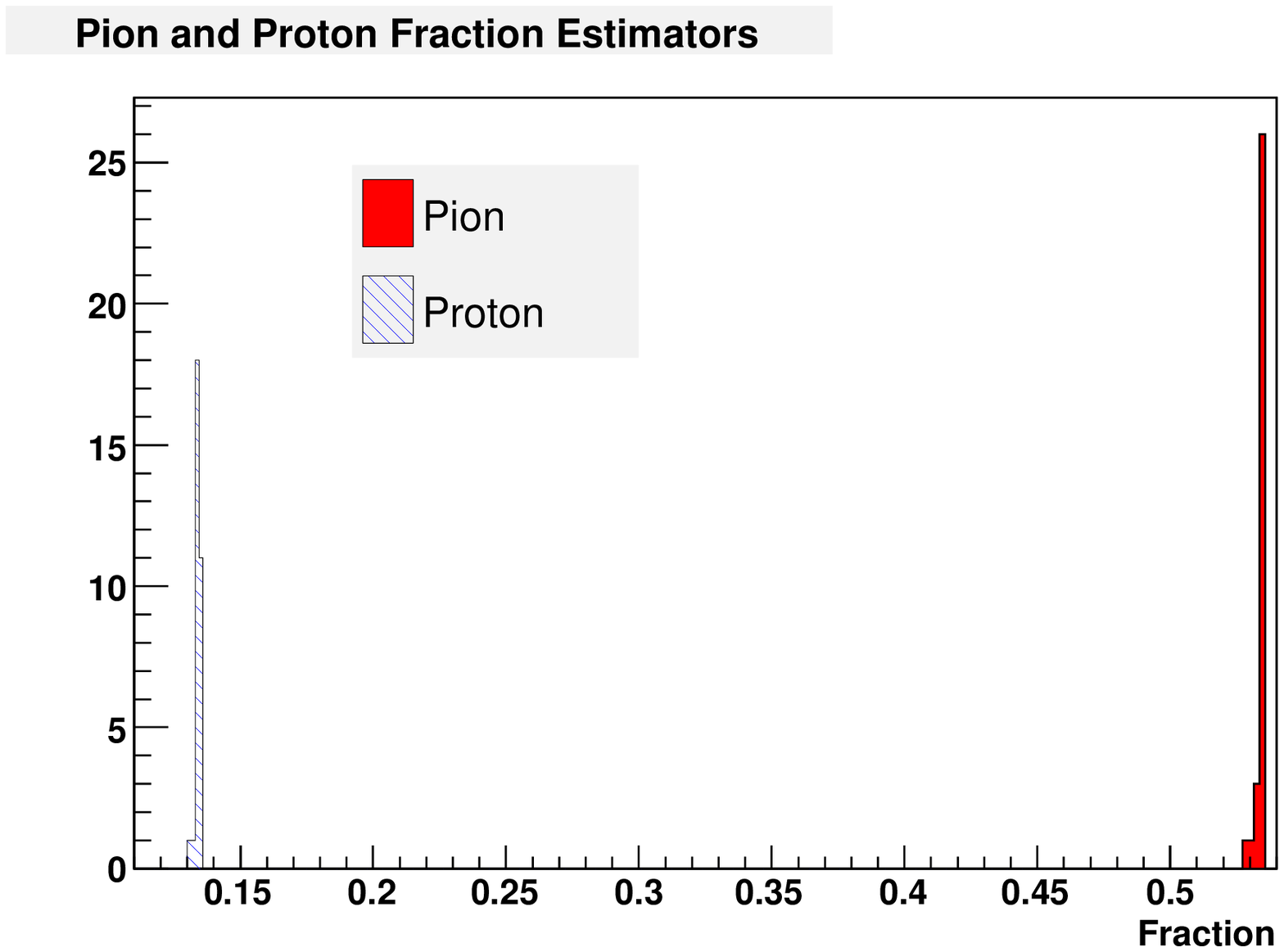,width=2.7in,height=1.4in}
\caption{The Pion and Proton fraction estimator distributions when the
complete (top) and incomplete (bottom) Likelihood expression is used.}
\label{fig:nobias}
\end{center}
\end{figure}

We  generated  a  sample   of  different  particle  types  with  known
composition as follow:
\begin{itemlist}
\item $PID$ variable is distributed, for each particle, according to a
typical  resolution  function (i.e.  the  template  used  in the  fit)
defined as:
\begin{equation}
 PID_{\mathrm{measured}} - PID_{\mathrm{expected}}(mom)
\label{eq:res}
\end{equation}
Note the dependence on momentum of the expected $PID$. 

It is important to note that we have chosen typical realistic values 
for all needed parameters.

This distribution represents:
\begin{equation}
 P(pid_{i} | mom_{i}, type_{j})
\label{eq:pidterm}
\end{equation}
in Eq. (\ref{eq:likeright}).
\item Momenta of the particles are distributed according a Gaussian 
$N(\mu_{j},\sigma_{j})$, where $j=\pi,K,P$ and:
\begin{equation}
 \nonumber{\mu_{\pi} = 1.00, \quad \mu_{K} = 1.25,\quad \mu_{P} = 1.25, }
\label{eq:mu}
\end{equation}
\begin{equation}
 \nonumber{\sigma_{\pi} = \sigma_{K} = \sigma_{P} = 0.50.} 
\label{eq:sigma}
\end{equation}
Those distributions obviously represent:
\begin{equation}
 P( mom_{i} | type_{j})
\label{eq:momterm}
\end{equation}
of equation \ref{eq:likeright}.
\item Particle fractions where fixed to:
\begin{equation}
 \nonumber{f_{\pi} = 50 \% ,\quad f_{K} = 35\% ,\quad f_{P} = 15\%.}
\label{eq:fractions}
\end{equation}
\end{itemlist}
We  then used  an  unbinned  Maximum Likelihood  fit  to estimate  the
particle  fractions  of  the  sample  using  the  Likelihood  function
described in Eq. (\ref{eq:likeright}) where:
\begin{equation}
 P(mom_{i} | type_{j}) = N(\mu_{j},\sigma_{j}).
\label{eq:gaus}
\end{equation}
In   Fig. \ref{fig:nobias} (upper  plot)   the  distribution   of  the
estimators for $f_{\pi}$ and $f_{P}$  are shown for thirty toy samples
of ten  thousand particles each.  As expected,  the fractions returned
by the fit are well centered on the true values given by the input.

Conversely,  the  same  distributions  obtained  with  the  incomplete
Likelihood function of Eq. (\ref{eq:likewrong}) (Fig. \ref{fig:nobias},
lower  plot) are  affected  by a  bias  much larger  than the  nominal
statistical uncertainty  of those measurements, due  to the difference
in the momentum distribution  of each particle type. This demonstrates
that the effect predicted in \cite{punzi} is actually very significant
in real--life problems of Particle Identification.
\subsection{The case of unknown momentum distributions}

Writing the complete  Likelihood function considering the distribution
of all  the observables used in  the fit is relatively straightforward 
in principle.

On the other  hand, in practice, we often  have poor information about
those distributions; sometimes they are completely unknown.  It is the
case,  for example,  of  the particle  fractions  produced during  the
fragmentation  of  heavy   quarks  where  the  corresponding  momentum
distributions are unknown and no functional hypothesis can be made.

Considering what was shown in  the previous section, we now wonder how
to avoid the bias and  write the complete Likelihood if the additional
observable distributions are unknown.

If no  specific functional form can be  assumed, we may want  to use a
general  one,  e.g.  we  could   consider  a  Series  Expansion  as  a
description of the distributions  with the expansion coefficients left
as free parameters to be determined by the fit.

We  then   write  the  momentum   term  of  the   Likelihood  function
(\ref{eq:likeright}):
\begin{eqnarray}
P(mom_{i}, type_{j})=\sum\limits_{m} a_{mj}U_{m}(mom_{i}) 
\label{eq:likeserie}
\end{eqnarray}
where $m$ is  the order and $U_{m}$ are the basis vectors used for the
series expansion.

Coming back to our toy sample, we considered Orthogonal Polynomials as
a  basis for the  expansion.  Amongst a  number  of possibilities,  we
selected Second Type Chebyshev Polynomials (denoted by $U_{m}$).

\begin{figure}
\begin{center}
\psfig{file=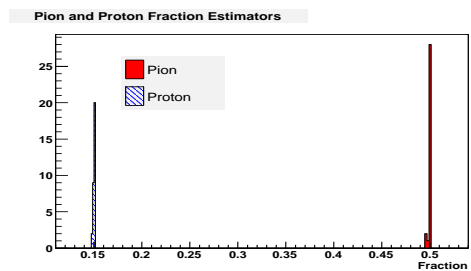,width=2.7in,height=1.4in}
\caption{The Pion and Proton  fraction estimator distributions using a
Series Expansion as a parameterization of the momentum distribution.}
\label{fig:serie}
\end{center}
\end{figure}

\begin{figure}
\begin{center}
\psfig{file=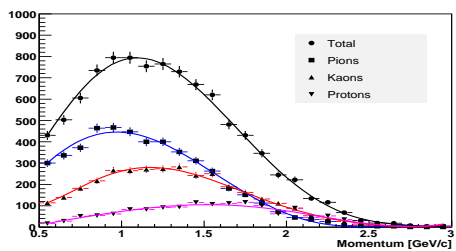,width=2.7in,height=1.4in}
\caption{The momentum projections  for each particle type superimposed
on the corresponding generated distributions.}
\label{fig:momenta}
\end{center}
\end{figure}

We   then  replaced  in   expression  (\ref{eq:likeright})   the  term
Eq.  (\ref{eq:gaus}) with  Eq. (\ref{eq:likeserie})  and  we performed
again the unbinned  Maximum Likelihood Fit, this time  by fitting also
the   parameters   of  the   polynomial   expansion.    As  shown   in
Fig. \ref{fig:serie}, now the bias is brought back to  zero, as it was
when  we   assumed  perfect  knowledge  of   the  individual  momentum
distributions of each particle type.
We have been  able to avoid the bias in the  fraction fit, without any
particular  assumption   on  the  functional  form   of  the  momentum
distributions.  In such a way we simulated the practical case where no
information  is known about  the additional  observable distributions.
Please notice also that just the  first seven terms of the Second Type
Chebyshev Expansion were needed  in order to parametrize each particle
type  momentum  distribution.   Another  interesting  aspect  is  that
comparing Fig. \ref{fig:serie} to Fig. \ref{fig:nobias} no significant
degradation in  the resolution of the estimator  is observed, although
the number of parameters is increased.   In Fig. \ref{fig:momenta} the
projections of the fit to the toy sample are shown.
\subsection{A more complicated case: Time of Flight}

\begin{figure}
\begin{center}
\psfig{file=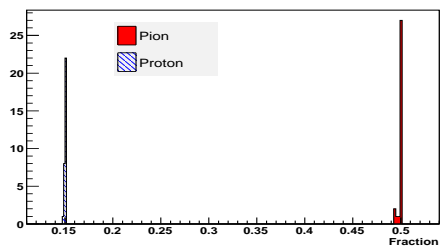,width=2.7in,height=1.4in}
\caption{The  Pion and Proton  fraction estimator  distributions using
two Series  Expansions as a  parameterization of the momentum  and the
arclength distributions.}
\label{fig:arc}
\end{center}
\end{figure}

\begin{figure}
\begin{center}
\psfig{file=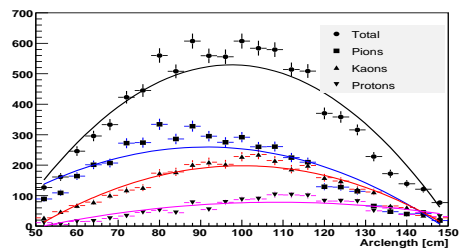,width=2.7in,height=1.4in}
\caption{The arclength projections for each particle type superimposed
on the corresponding generated distributions.}
\label{fig:arcproj}
\end{center}
\end{figure}

Suppose that our  $PID$ information is obtained by  the measurement of
the Time of  Flight. The expression of the expected  TOF is a function
of two observables:
\begin{eqnarray}
TOF_{{\mathrm{expected}}}(mom,L) =\frac{L/c}{\sqrt{1+{(m_{j}/mom)}^{2}}}
\label{eq:exptof}
\end{eqnarray} 
where  $L$ is the  length travelled  by the  particle during  its time
measurement (arclength) and it is a functon of the production angle of
the particle  (in  the cylindrical geometry of the  TOF detector), $c$
is the speed of light, $m_{j}$  is the mass of the particle hypothesis
$j$  and  $mom$ is  again  the momentum.  Both  the  momentum and  the
arclength  distributions could  be different  for each  particle type,
i.e.,  both  observables could  be  source  of  bias in  the  particle
fractions estimation.   Assuming no correlations  between the momentum
and   the    arclength,   we    have   to   modify    the   expression
(\ref{eq:likeright}) to be:
\begin{eqnarray}
L(f_{j})=\prod\limits_{i}(\sum\limits_{j=\pi,K,P}    f_{j}   P(pid_{i}
mom_{i},                                                       arc_{i}|
type_{j})\\\nonumber{=\prod\limits_{i}(\sum\limits_{j=\pi,K,P}    f_{j}
P(pid_{i}   |   mom_{i},   type_{j})}\\\nonumber{\times  P(mom_{i}   |
type_{j})}\\\nonumber{\times P(arc_{i} | type_{j})).}
\label{eq:likearc}
\end{eqnarray}
We  then added  the  simulation of  the  arclength in  our toy  sample
according to a normal distribution $N(\mu_{j},\sigma_{j})$ using  the 
values:
\begin{equation}
 \nonumber{\mu_{\pi} = 90, \quad \mu_{K} = 100,\quad \mu_{P} = 110, }
\label{eq:muarc}
\end{equation}
\begin{equation}
 \nonumber{\sigma_{\pi} = \sigma_{K} = \sigma_{P} = 25.} 
\label{eq:sigmaarc}
\end{equation}
Considering again the case where no information is available about the
distributions of each particle type, we used the same technique of the
Series Expansion  for both variables.   We repeated our fit  on thirty
toy samples and also in this case, as  shown in Fig. \ref{fig:arc}, no
bias was observed for our estimator. It is also interesting to observe
that  we used  just three  terms of  the Chebyshev  Expansion  for the
arclength   parameterization,   that   results  in   an    approximate
description     of    data     (see    arclength     projections    in
Fig. \ref{fig:arcproj}) but it doesn't affect the results of the fit.
\vspace*{-12pt}

\section{Conclusions}

In this  short paper we focused  on a practical and  common problem of
particle  physics: the estimation of the particle type fractions using
Particle Identification information. We showed that a significant bias
can arise from  the use of an incomplete  expression of the Likelihood
under realistic  conditions.  We  also considered a  practical problem
where no  information was assumed about an  observable.  We eliminated
the bias  by using Series  Expansions of the unknown  distributions in
orthogonal polynomials,  where the coefficients of  the expansions are
free  parameters determined  by the  fit.  We  also considered  a more
complicated  example  where  two  relevant  observables  have  unknown
distributions,  and  also  in  this  case  the  Series  Expansion  was
successful in avoiding  biases in  determining the  fractions  of each
component.
 
\balance

\end{document}